# Two Dimensional heterostructure and its application in efficient quantum energy storage

Meenakshi Talukdar, Sushant Kumar Behera and Pritam Deb*

*Advanced Functional Material Laboratory (AFML), Department of Physics, Tezpur University (Central University), Tezpur-784028, India.*

*Corresponding author

Email address: pdeb@tezu.ernet.in (Pritam Deb)

Portable miniaturized energy storage micro-supercapacitor has engrossed significant attention due to its power source and energy storage capacity, replacing batteries in ultra-small electronic devices. Fabrication with porous and 2D graphitic nanomaterials with high conductivity and surface area signify high performance of micro-supercapacitor. In order to satisfy the fast-growing energy demands for the next-generation, we report performance and design of a 2D heterostructure of EDLC (g-$C_3N_4$) & pseudocapacitive ($FeNi_3$) resulting low ionic diffusion path and prominent charge storage based on their synergic functionalities. This heterostructure system shows an enhanced quantum capacitance (38% enhancement) due to delocalized states near Fermi level. Having achieved the areal capacitance of 19.21 mFcm$^{-2}$, capacitive retention (94%), enhanced power density (17 fold) having ultrahigh energy density of 0.30 Wh.cm$^{-3}$ and stability of the material even without any obvious degradation after 1000 cycles, this smart heterostructure acts as a new platform for designing high-performance in-plane micro-supercapacitor.

Rate of electric energy consumption has dramatically increased due to growing global population and fast development of modern industries. There is an unmet need of high-performance energy storing systems, where large amount of energy need to be delivered or accepted within short period of time[1]. Micro-supercapacitors are promising energy storage device for future due to their high energy storage within small solid structure[2]. Compared to other batteries, in-plane designed micro-supercapacitors are preferred for developing electronic devices due to its unique properties in excellent rate capability, flexibility[3], stability and durable cycle life[4,5]. As battery performance has some limitations, in-plane micro-capacitors are the potential candidates to store the energy in miniaturized portable form directly on a chip meeting the demands and ensuring the adequate[6,7]. Electrochemical capacitors (EC) are attracting much attention towards storage of energy by redox reaction or by ion adsorption. EC acts as mediator between the energy and power electrostatic capacitors which works mainly on electrochemical double layer (EDL) and pseudo-capacitive materials. EDL follows ion adsorption, delivering higher power while pseudo-capacitive delivers higher energy density[8] suffering from redox reaction at the surface of the electrode. Future generation demands portable energy power source where supercapacitors with thinner, small sized and highly flexible[9] as similar to micro-batteries will be required. In this case, designed and smart nanomaterials need to be developed which will act as storage source in the form of in-plane micro-supercapacitors overcoming the natural limits such as poor conductivity[10], instability and maximizing their significant advantages.

In this aspect, two dimensional (2D) materials with large surface area and abundance of reaction sites, shows high possibility of large charge storage capacitance [11]and better performances in terms of stability and flexibility. 2D materials also exhibit unique properties such as high power density, excellent rate performance and long cycle life[12], pushing the energy storage to the next level of stored specific charges. Owing to strong charge carrier localization, the ionic conductivity[13] in two dimensional layered materials is efficient throughout the surface rather than between the layers[14]. Pseudocapacitive materials with 2D material in composite will exhibit larger specific capacitances, whereas double layer type materials exhibit better rate handling capability[15] and superior cycle longevity. In recent times, heterostructure have triggered great scientific and technological interests due to their high versatility as the essential components in quantum devices. Moreover, 2D heterostructure with efficient manipulation of physical property opens application possibilities in broad new range of nano-electronic devices. Apart from its high surface area, the presence of mesopores within the sheet[16] helps in sufficient transportation of electrolyte shortening the diffusion distance. However, pseudocapacitive material with fast reversible surface redox properties is required to improve the energy storage of EDLC materials. Iron based materials are promising materials for electrochemical devices because of their low cost, nontoxicity[17,18], good chemical stability and high energy-power density[19]. However, the materials suffer from aggregation after reaction, poor capacity retention and low electronic conductivity which hinder the electrolyte ions accessibility. Loading of nanoparticles on graphitic sheets can not only overcome these problems[20], but also minimize the aggregation or restacking of the nanosheets. Presence of high content of nitrogen in the 2D graphitic sheet can stabilize pseudocapacitive $FeNi_3$ redox material by escalating the metal-carbon binding energy and increases the capacitive performances. In this aspect, we have developed a novel electrode of 2D mesoporous heterostructure material as an in-plane micro-supercapacitor for energy storage. The electrode material is based on mesoporous carbon nitride embedded with redox $FeNi_3$ nanoparticles showing high areal capacitance performance. Presence of nitride content and mesopores in the two dimensional sheets provide more reaction sites for metal electrode allowing the electrolyte to efficiently exploit the surface. Also synergic combination of both 2D carbon nitride and $FeNi_3$ nanoparticles will not only improve the physical properties, but also influences the rate of intercalation within the system due to continuous conducting pathway in carbon nitride. In this aspect, 2D g-$C_3N_4$ hybridizes with pseudocapacitive material for enhancement in electrochemical performances.

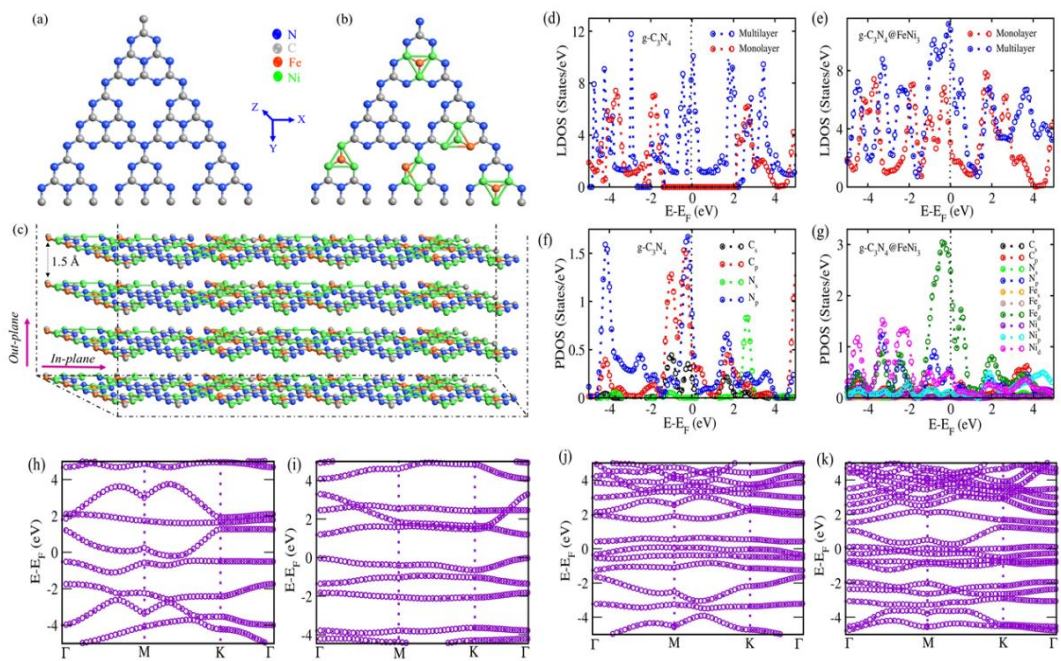

**Figure 1│ Optimized geometry**. a) g-$C_3N_4$ and b) g-$C_3N_4$@FeNi$_3$ heterostructure from the top view; c) Multi-layered heterostructure system with an interlayer spacing of 1.5 Å in a supercell of 3×3×1 units marked with black dotted line. Pink coloured arrows marks are shown to distinct out-plane and in-plane direction of the supercell model for easy understanding of the calculation method with respect to the Cartesian coordinate system. Local density of states (LDOS) calculation for both d) g-$C_3N_4$ and e) g-$C_3N_4$@FeNi$_3$ heterostructure system. In both cases, monolayer and multi-layered sheets are considered. Similarly, partial density of states (PDOS) is shown for f) g-$C_3N_4$ and, g) g-$C_3N_4$@FeNi$_3$ heterostructure system. Fermi energy level is shown with black dotted line. Electronic band structure of both monolayers h) g-$C_3N_4$ and i) g-$C_3N_4$@FeNi$_3$ heterostructure system are shown. Similarly, the band structure of respective multilayer systems is shown in j) g-$C_3N_4$ and k) g-$C_3N_4$@FeNi$_3$ heterostructure. The bands are plotted along the symmetric k-points of Γ-M-K-Γ direction.

The minimum energy sustained geometry of the heterostructure system along with pristine form of graphitic sheets is fixed via optimization of surface atomic configuration. The optimization is followed by moving the atoms assisted with the structure to the minimum energy positions for accurate stability during computation. The stable geometry of the systems is shown in (Fig. 1(a) to (c)). The geometry and corresponding interlayer distances are determined by employing the atomic structure optimization with the van der Waals (vdW) corrections (Table S2 for stable atomic coordinates and Table S3 for various energy components of the stable geometry) [21]. It is observed that presence of metallic atoms (i.e. Fe and Ni) on the graphitic sheet surface make the heterostructure system stable at higher total surface energy compared to its pristine sheet (Table S3). The stable atomic configuration of g-$C_3N_4$ (shown in Fig. 1 (a)) sheet achieved the minimum energy stability at -768.83 Ry, whereas the monolayer heterostructure (g-$C_3N_4$@$FeNi_3$) (Fig. 1 (b)) showed energy value of -1068.09 Ry. In supercell structure (Fig. 1 (c)), the stable energy has achieved a value of -964.25 Ry. The enhancement in surface free energy value in heterostructure system controls the electronic structure and makes them dynamic towards the molecular motion with enhanced kinetic energy compared to pristine system.

Local densities of states (LDOS) (Fig. 1 (d) and (e)) have been calculated to understand the origin and control of molecular motion for energy storage activity of the active sites formed in g-$C_3N_4$@$FeNi_3$ heterostructure system. Overlapping states are observed from the plots of the density of states (Fig. 1) showing the dynamic and metallic behaviour in the monolayer and multilayer systems compared to their pristine sheet. Fe and Ni atoms presence makes the host metallic in nature with sufficiently large states near Fermi level (Supplementary Fig. S6 and S7). Presence of localized electrons due to embedded metal atoms (Fe and Ni atoms) over g-$C_3N_4$ surface contributed towards the formation of overlapping states at Fermi level within conduction band and improved conductivities with delocalized nature of the active metal sites (Supplementary Fig. S8 and S9). Carbon and nitrogen atom orbitals show significant states which are delocalized in heterostructure system unlike their localized nature in pristine g-$C_3N_4$ sheet (Supplementary Fig. S11). Moreover, this delocalized electronic behaviour supports in achieving highly negative surface Gibbs free energy value (Supplementary Fig. S13). This nature helps in maintaining the metallic behaviour (Supplementary Fig. S11 and S12) of the heterostructure system for easy energy storage.

To realize the orbital existence and contribution, the DOS pattern has been projected in reciprocal *k*-space to get projected crystal wave function values named as projected density of states (PDOS). Here, we have resolved the case for both pristine and heterostructure system to show the PDOS (shown in Fig. 1 (f) and (g)) pattern. The PDOS pattern clearly indicates the reflection of all orbitals of the constituent atoms (i.e. carbon, nitrogen, iron and nickel) in monolayer and multi-layer pattern of the systems. It is noticed that there is no enhancement in projection densities of the systems in both monolayer and multilayer case. This confirms equal contribution and overall projection of all atomic orbitals that are present in the pristine and heterostructure systems. This PDOS pattern reflects the orbital splitting which can be corroborated with respective electronic band structure in (Fig. 1 (h) to (k)). We have noticed that in the presence of Fe and Ni atoms on g-$C_3N_4$ surface, the band gap is completely closing indicating an overlapping metallic stage. This can be attributed to the effect of metal atoms on graphitic sheet substrate to retain the metallic nature in the heterostructure system.

Microstructure and morphology of the prepared 2D heterostructure nanomaterial for in-plane supercapacitor shown in (Fig. 2) was characterized by HRTEM and FESEM analysis. (Fig. 2a, b) shows the individual TEM images of graphitic carbon nitride and g-$C_3N_4$@$FeNi_3$ heterostructure material. (Fig. 2a) confirms the formation of mesopores in the sheets of carbon nitride. Advantage of mesopores in case of storage application is to create more channels for electrolyte ions to transport electrochemically[22], accessing more electroactive sites for energy storage faster. Also, porous materials act as the storage of ions and electrons resulting in improving the mass and charge transfer throughout the surface (Supplementary Fig. S4). (Fig. 2b) signifies the dispersion of homogeneous $FeNi_3$ nanoparticles with typical diameters 100-200 nm on the graphitic sheets with less aggregation (Supplementary Fig. S3). The uniformity and less aggregation will enable enhancement of the electrochemical property (shows in Supplementary Fig. S5). FESEM images showed in (Fig. 2c) reveals the formation of stacked graphitic sheets with spherical shaped $FeNi_3$ nanoparticles over the sheets. The formation of heterostructure g-C$_3$N$_4$@FeNi$_3$ system can be further justified from the given (Supplementary Fig. S2) microstructural images.

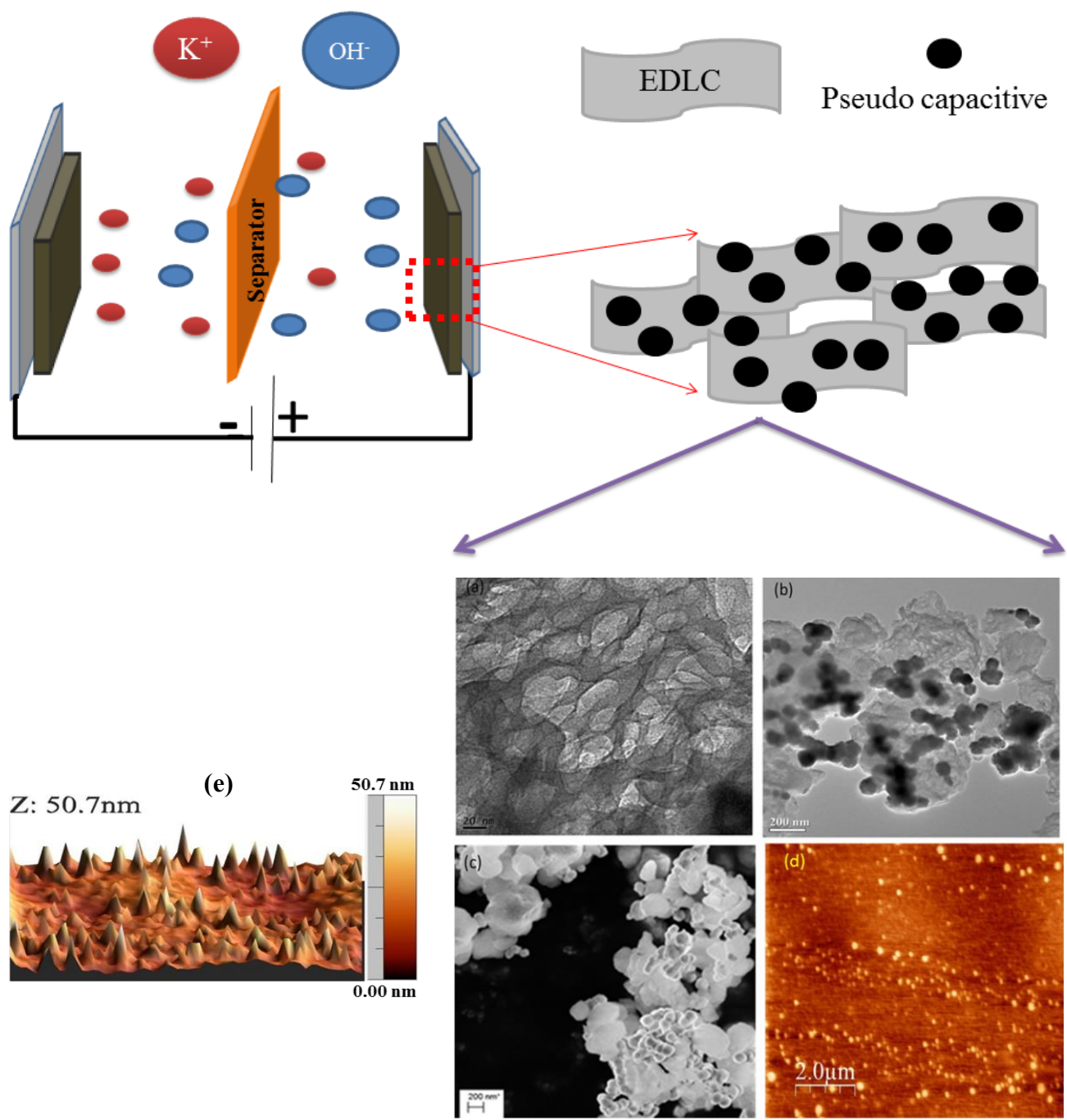

**Figure 2 │ Schematic and microstructural characterization 2D heterostructure for in-plane micro-supercapacitor**; a)TEM micrograph of g-C$_3$N$_4$; b) g-C$_3$N$_4$@FeNi$_3$ heterostructure; c)FESEM image of g-C$_3$N$_4$@FeNi$_3$ ; d) and e) AFM images of g-C$_3$N$_4$@FeNi$_3$.

AFM images of the system are shown in (Fig. 2d, e) respectively for evaluating surface topology and roughness. The images reveal the thickness of the prepared multilayer g-$C_3N_4$@$FeNi_3$ heterosystem to be 50 nm. Justifying with the AFM results, (Fig. 2 (b)) further signifies the formation of multilayer heterostructure system, where $FeNi_3$ is localised on the surface of g-$C_3N_4$ layers. Raman Analysis of g-$C_3N_4$@$FeNi_3$ (Supplementary Fig. S1) displays two peaks at 804 and 1782 cm$^{-1}$, assignable as G band and D band marks of graphitic carbon materials respectively. D band peak at 804 cm$^{-1}$ corresponds to disordered carbon, which can be attributed to the defects and partially disordered structure of graphitic layers in the heterosystem. The broad bands at around 2429 cm$^{-1}$ are characteristic of 2D graphitic carbon. Also the decrease in ratio of $I_D/I_G$ (ratio as 0.76) indicates defect free in the heterostructure system.

Cyclic voltammetry studies and galvanostatic charge-discharge studies have been carried out in a three electrode system taking Ag/AgCl and Pt as the reference and counter electrode. The used electrolyte is KOH with two different concentrations (1M and 2M). (Fig. 3a) reveals the cyclic voltammetry curves of the prepared g-$C_3N_4$@$FeNi_3$ heterosystem under various scan rates (25-200 mVs$^{-1}$). The curves remained unchanged, stable and nearly rectangular in shape indicating high flexibility and good electrochemical stability. Also the quasi rectangular profile exhibiting symmetry in shape, suggests an ideal capacitive behaviour of prepared heterostructure nanomaterial electrode. Such quasi-rectangular shape indicates that the mechanism of charge storage is due to electric double layer capacitance (EDLC) and Faradic reactions (pseudo capacitance) within the voltage range (-0.2V to + 0.6V). Moreover, redox peaks are completely absent during the CV scan which is due to the formation of double layer at the electrode and electrolyte interfaces. Because of the presence of $OH^-$ ions in the electrolyte, there is a slight deviation in the curves indicating the reversible reaction. (Fig. 3d and 3e) show the galvanostatic charge discharge (GCD) curve under various current densities with varying molar electrolyte concentration (KOH) at 1M and 2M. The discharge areal capacitance has been calculated using the formula, $C_a = \frac{I*t}{S*\Delta V}$, where $I$ is the applied current (mA), $t$ is the time of discharge (s), $S$ is the area of working electrode (cm$^{-2}$), and $\Delta V$ is the potential window. During GCD curve, there is a voltage plateau observed in the first curve of 1M KOH electrolyte solution. The plateau is observed as a consequence of ion intercalation. This process creates interlayer spacing and expansion within the sheets leading to swelling of the electrode. Besides, it has been observed that increment in current density value results in reducing the plateau. Also, no straight voltage curve and height changing curves are seen in the GCD curve. It implies that the mechanism behind charge storage[23] is a mixture of both intercalation and adsorption. The areal capacitances are 5.12 mFcm$^{-2}$, 4.71 mFcm$^{-2}$ and 4.28 mFcm$^{-2}$ for 0.05 mA cm$^{-2}$, 0.15 mA cm$^{-2}$ and 0.24 mA cm$^{-2}$ current densities respectively, for electrolyte concentration of 1M KOH. Ion intercalation depends on the electrolyte concentration[24]. The ion intercalation decreases as we increase the electrolyte concentration to 2M KOH. On increasing the concentration of electrolyte, there is a formation of anion film above the surface of the sheets due to which no voltage plateau is observed in (Fig. 3e). As a result there is an increase in the areal capacitance of 19.21 mFcm$^{-2}$, 17.32 mFcm$^{-2}$ and 8.180 mFcm$^{-2}$ for 0.05 mA cm$^{-2}$, 0.15 mA cm$^{-2}$ and 0.24 mA cm$^{-2}$ current densities respectively, leading towards change in kinetics of the system. The reason behind the absence of plateau is due to the change or aggregation of dominant structure in the electrolyte which results in decrease in the potential of the superior electrolyte concentration[25]. Symmetry and linear curves at high current density indicates the excellent reversibility [26] during charge-discharge process. Limited migration of ions in electrode creates difficulty in charging process exhibiting decrease in specific capacitance on further increasing in current density. Higher concentration of electrolyte promotes decreasing oxidation and reduction charges. At higher concentration of electrolyte, it was easier for the ions to transport within the electrode layers, leading to an effective building-up for double layer and hence increases the specific capacitance. However, if the concentration of electrolyte was low (i.e. 1M KOH), the ions inside the pores decrease the diffusion current, moreover it provided an insufficient number of ions for double-layer to build-up[27,28].

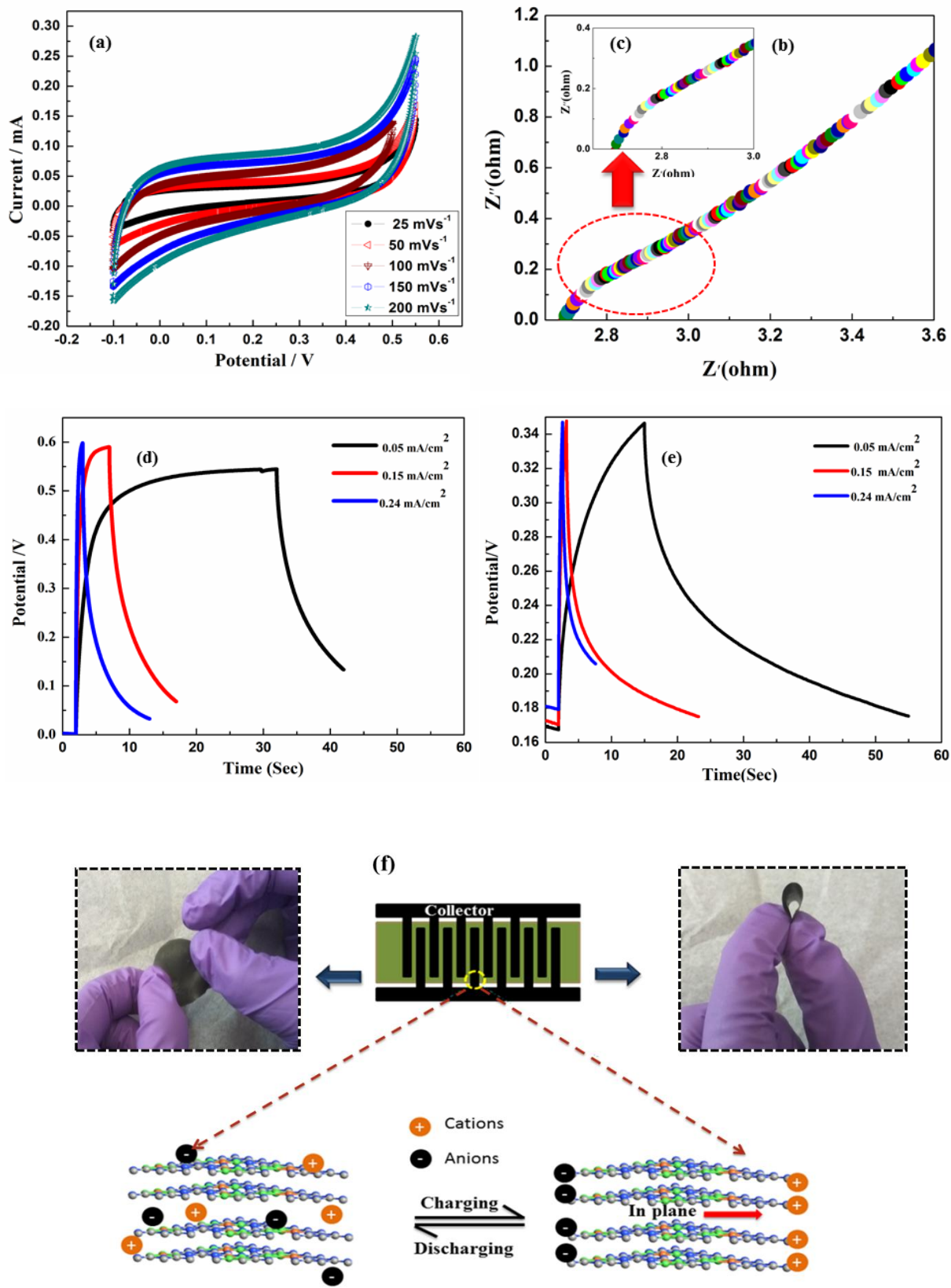

**Figure 3 | Electrochemical performance** a) g-$C_3N_4$@$FeNi_3$ heterostructure samples under three-electrode mode at different scan rates; b) Nyquist plots of the electrodes and c) inset shows the enlarged high-frequency region; Galvanostatic charge discharge curve of the prepared heterostructure system d) and e) at different electrolyte concentration; f) Mechanism for the device fabrication using 2D g-$C_3N_4$@$FeNi_3$ heterostructure.

Nyquist plot as shown in (Fig. 3b), of the as prepared heterostructure material has been carried out under the frequency range of 0.30 MHz to 20 Hz, inset to the figure (Fig. 3c) showing the expanded high-frequency region of the same plot. In the curve, two distinct parts i.e. a small semicircle arc at high frequency region and a linear part at low frequency region can be seen. The vertical line at low-frequency region indicates the good capacitive behaviour and low diffusion resistance of the heterostructure nanomaterial. It is eminent that the Nyquist curve of the g-$C_3N_4$@$FeNi_3$ heterostructure electrode exhibits very small arc and straight line, indicating ideal capacitive behaviour of the device and low resistance of the electrodes. The higher capacitance of electrode is due to the combined heterostructure as well as the synergistic effect of $FeNi_3$ nanoparticles along with the porous 2D graphitic sheets. The semi-circular loop at high frequency is too small to identify, indicating the small interfacial charge transfer resistance and good electrical conductivity [29-30] of the electrode material which enables improved electronic and ionic conduction of the g-$C_3N_4$@$FeNi_3$ heterostructure.

We demonstrated an effective strategy for the development of an electrode using g-$C_3N_4$@$FeNi_3$ heterostructure. Graphitic carbon nitride with highly mesoporous nature effectively enhanced the transmission of ions and electrons simultaneously $FeNi_3$ nanoparticles prevented the aggregation of graphitic layers by increasing the areal capacity of the system. Complete electronic structure and quantum capacitance of the heterostructure system have been investigated systematically. The DFT simulations show that the heterostructure generates new delocalized electronic states, consequently enhance overlapping states in band structure near Fermi level and derive 38% enhancement in quantum capacitance. The synergic effect of both electro double layer and pseudocapacitive components provide several electron channels for ion transfer within the system. The powder density (with super high energy density of 0.30 Wh.cm$^{-3}$) of this system is approximately 17 times higher than that of conventional available supercapacitor. Apart from high areal capacitance, capacitive retention, excellent stability and flexibility, the unique strategy of this 2D architecture heterostructure system is the long duration in discharge time which occurred due to conductivity enhancement in ion transport, makes the system a promising candidate for ultrahigh performance in-plane micro-supercapacitor for future generation.